\documentclass[aps,prd,reprint,amsmath,amssymb,longbibliography,nofootinbib]{revtex4-2}
\usepackage{graphicx}
\usepackage{bm}
\usepackage{amsmath}
\usepackage{amsfonts}
\usepackage{amssymb}
\usepackage{color}
\usepackage{hyperref}
\usepackage{lipsum}
\usepackage{dcolumn}
\usepackage[dvipsnames]{xcolor}

\usepackage[capitalise]{cleveref}
\usepackage[activate={true,nocompatibility},final,kerning=true,factor=1100,stretch=10,shrink=10]{microtype}
\usepackage{academicons}
\usepackage{fontawesome5}
\definecolor{orcidlogocol}{rgb}{0.65, 0.807, 0.223}
\newcommand{\orcid}[1]{$\,$\href{https://orcid.org/#1}{\textcolor{orcidlogocol}{\faOrcid}}}

\newcommand{\Planck}{\textit{Planck}}

\newcommand{\lcdm}{$\Lambda$CDM}

\newcommand{\beq}{\begin{equation}}
\newcommand{\eeq}{\end{equation}}
\renewcommand{\arraystretch}{1.6}

\def\ee{\end{equation}}
\def\bea{\begin{eqnarray}}
\def\eea{\end{eqnarray}}
\def\bse{\begin{subequations}}

\begin{document}
\title{A Fast and Accurate Implementation of the Effective Fluid Approximation for Ultralight Axions}
\author{Adam Moss\orcid{0000-0002-7245-7670}}
\email{adam.moss@nottingham.ac.uk}
\affiliation{University of Nottingham}
\author{Lauren Gaughan \orcid{0009-0004-8338-0180}}
\email{lauren.gaughan@nottingham.ac.uk}
\affiliation{University of Nottingham}
\author{Anne M. Green\orcid{0000-0002-7135-1671}}
\email{anne.green@nottingham.ac.uk}
\affiliation{University of Nottingham}
\date{\today}

\begin{abstract}
We present a numerically efficient and accurate implementation of the Passaglia-Hu effective fluid approximation for ultralight axions (ULAs) within the Boltzmann code CAMB. This method is specifically designed to evolve the axion field accurately across cosmological timescales, mitigating the challenges associated with its rapid oscillations. Our implementation is based on the latest version of CAMB, ensuring compatibility with other cosmological codes., e.g. for calculating cosmological parameter constraints. Compared to exact solutions of the Klein-Gordon equation, our method achieves sub-percent accuracy in the CMB power spectrum across a broad range of axion masses, from $10^{-28}\,\mathrm{eV}$ to $10^{-24}\,\mathrm{eV}$. We perform Markov Chain Monte Carlo (MCMC) analyses incorporating our implementation, and find improved constraints on the axion mass and abundance compared to previous, simpler fluid-based approximations. For example, using \Planck\ PR4 and DESI BAO data, we find $2\sigma$ upper limits on the axion fraction $f_{\rm ax} < 0.0082$ and physical density $\Omega_{\rm ax}h^2 < 0.0010$ for $m=10^{-28}$ eV. The code is publicly available at \url{https://github.com/adammoss/AxiCAMB}.
\end{abstract}

\maketitle

\section{Introduction}

The \lcdm\ paradigm provides a remarkably successful framework for describing the evolution of the Universe \cite{Planck:2018vyg}. However, the precise nature of the dark sector remains a central open question, motivating the exploration of a wide array of dark matter \cite{Cirelli:2024ssz} and dark energy \cite{Copeland:2006wr} candidates. Among these, ultralight axions (ULAs) \cite{Svrcek:2006yi,Arvanitaki:2009fg,Marsh:2015xka} have attracted significant attention due to their potential to address persistent cosmological puzzles, including anomalies observed in small-scale structure \cite{Hu:2000ke} and the fundamental nature of dark energy \cite{Frieman:1995pm}. While initially proposed to resolve the strong-$CP$ problem in quantum chromodynamics \cite{Peccei:1977hh,Weinberg:1977ma,Wilczek:1977pj}, axions with masses considerably smaller than typical QCD axions are predicted in various extensions of the Standard Model, such as string theory compactifications \cite{Svrcek:2006yi,Arvanitaki:2009fg}.

The cosmological behavior of ULAs is strongly influenced by the relationship between their mass, $m$, and the Hubble parameter, $H$ \cite{Hlozek:2014lca}. When $H > m$, the axion field effectively remains constant due to Hubble friction, behaving like vacuum energy. As the universe expands and the Hubble rate decreases below the axion mass ($H < m$), the field begins to oscillate. In this oscillatory regime, the axion energy density redshifts like cold dark matter (CDM). This transition, which occurs around the time the Hubble rate is comparable to the axion mass, introduces a distinct cosmological signature, particularly for axion masses close to the value of the Hubble parameter at matter-radiation equality, $H_{\text{eq}} \sim 10^{-28}\,\mathrm{eV}$. Consequently, axions with masses close to this scale can introduce unique dynamics and observable effects.

\begin{figure}
    \includegraphics[width=\columnwidth]{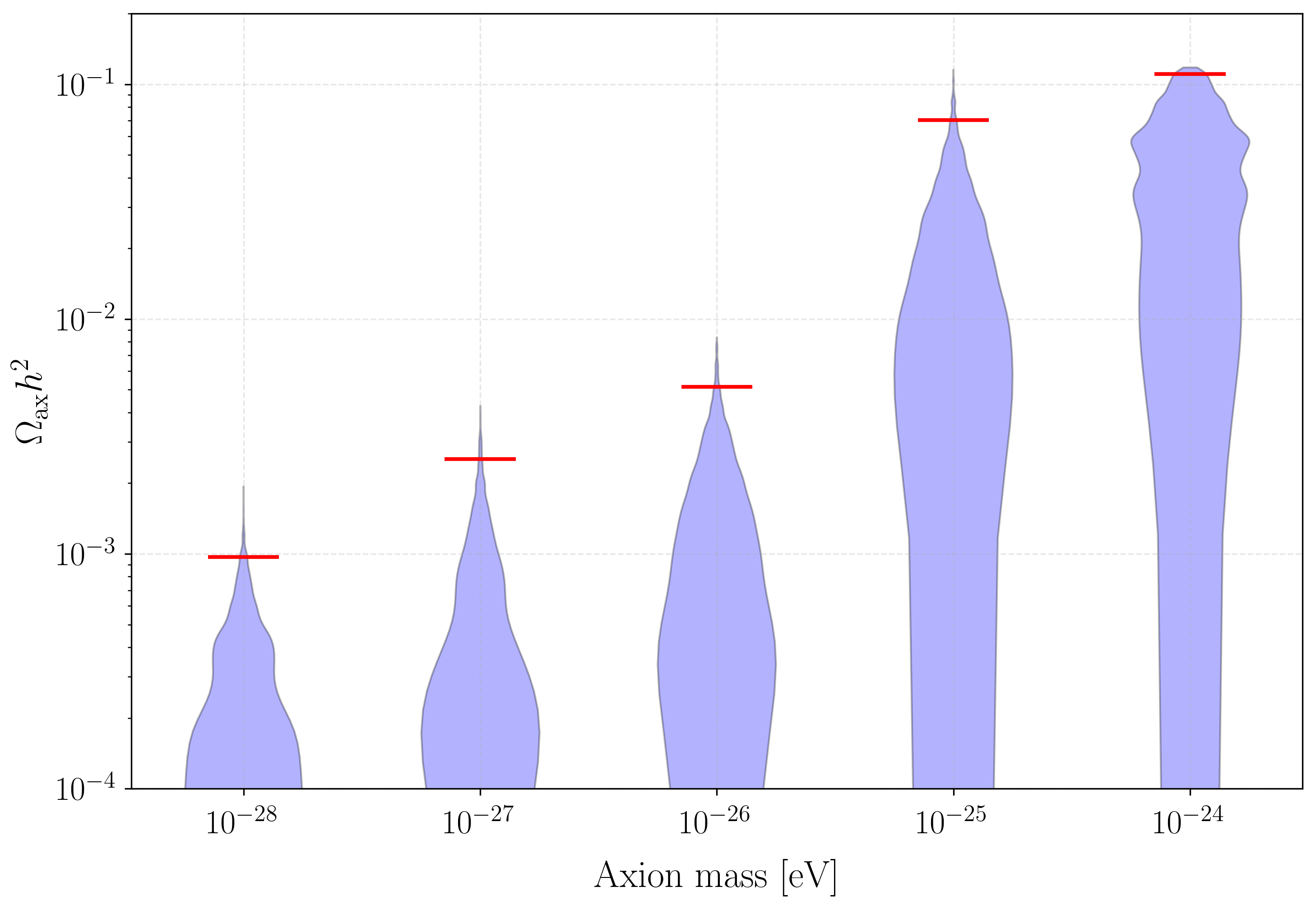}
\caption{Violin plots of the posterior distributions for the axion density, 
$\Omega_{\mathrm{ax}}\,h^2$, at fixed axion masses, $m\in\{10^{-28},10^{-27},10^{-26},10^{-25},10^{-24}\}\,\mathrm{eV}$. The violin shapes illustrate the full 
marginalized distribution, while the horizontal red bars indicate 
the $2\sigma$ upper limits. These results are derived using our implementation of the Passaglia--Hu 
EFA combined with \Planck\ PR4 data~\cite{rosenberg22} and 
DESI BAO~\cite{DESI:2024mwx}.}
    \label{fig:m_axion_vs_f_axion}
\end{figure}

One notable observational consequence of ULAs is the suppression of structure formation on small scales \cite{Hu:2000ke}. For this reason, axions in the mass range $10^{-28} \, {\rm eV} \lesssim m_{\text{ax}} \lesssim 10^{-25}$ eV have been shown (e.g. Ref.~\cite{Rogers:2023ezo}) to reduce the $S_{8}$ tension (for a review see e.g. Ref.~\cite{Abdalla:2022yfr}), as they suppress structure growth on $\sim 8h^{-1}\,\mathrm{Mpc}$ scales. Heavier axions, with masses around $m \sim 10^{-22}\,\mathrm{eV}$, are expected to damp density fluctuations on kpc scales, a scenario often referred to as ``fuzzy dark matter''. However, direct numerical simulation of the axion field evolution, in particular for masses much smaller than the Hubble parameter today ($m \ll H_0 \sim 10^{-33} \, \mathrm{eV}$), becomes computationally prohibitive. This is because the oscillation timescale of the field becomes much shorter than relevant cosmological timescales, requiring very small time steps. This poses a significant challenge for cosmological Boltzmann codes, which aim to accurately model the evolution of perturbations across a wide range of scales.

A common approach to circumvent these numerical difficulties is the \emph{effective fluid approximation} (EFA) \cite{Hu:2000ke,Hlozek:2014lca}. By treating the axion field as an effective fluid, one can average over the rapid field oscillations, resulting in simplified equations of state and sound speeds. However, the standard EFA can introduce systematic errors \cite{Cookmeyer:2019rna}, which are particularly significant for axion masses in the range $10^{-28}\,\mathrm{eV}\lesssim m\lesssim 10^{-26}\,\mathrm{eV}$, where current Cosmic Microwave Background (CMB) data already provide strong constraints from the power spectrum. Furthermore, even for heavier axions ($10^{-24}\,\mathrm{eV} \lesssim m \lesssim 10^{-19}\,\mathrm{eV}$), overly simplified EFAs can lead to inaccurate characterizations of their cosmological effects on very small scales. An alternative cycle-averaging method was developed by Ure\~na-L\'opez and Gonzalez-Morales \cite{Urena-Lopez:2015gur}, however, Ref.~\cite{Cookmeyer:2019rna} showed that it is  equivalent to the standard EFA.

Current cosmological constraints on ULA utilise the standard EFA approach implemented in codes such as \texttt{AxionCAMB} \cite{Hlozek:2014lca,Hlozek:2017zzf} and \texttt{AxiCLASS} \cite{Poulin:2018dzj}. Using CMB data from the \Planck\ satellite, Atacama Cosmology Telescope (ACT) and South Pole Telescope (SPT), and Baryon Oscillation
Spectroscopic Survey (BOSS) galaxy clustering data, Ref.~\cite{ Rogers:2023ezo} found,
using \texttt{AxionCAMB}, that the axion dark matter fraction, $f_{\text ax}$, is $<0.1$ for $m_{\text{ax}} \lesssim 10^{-26}$ eV and $<0.01$ for $10^{-30} \, {\rm eV} \lesssim m_{\text{ax}} \lesssim 10^{-28}$. Ref.~\cite{Winch:2023qzl} presented a method for modelling so-called `extreme' axions, where the initial value of the axion field is sufficiently large that the full cosine potential is important, which combines the generalized dark matter model with the EFA. For extreme axions structure formation on small scales is enhanced rather than suppressed, and they find that constraints on the axion fraction can be significantly weakened. With the approach of CMB-S4 \cite{Abazajian:2019eic} and other more precise cosmological datasets, developing fast and accurate methods for calculating axion density perturbations is important.

To address the limitations of the standard EFA, Passaglia and Hu~\cite{Passaglia:2022bcr} developed an improved approximation scheme. Their method systematically removes the rapid oscillations by introducing auxiliary fields, and then calibrates the effective fluid equation of state and sound speed against exact solutions of the Klein-Gordon equation. This approach maintains accuracy throughout the field's evolution, reducing errors across a wide range of axion masses. Importantly, this refined prescription allows for a more reliable incorporation of ULAs into standard cosmological analyses without incurring excessive computational cost.

In this paper, we detail our implementation of the improved EFA developed by Passaglia and Hu within \texttt{CAMB}~\cite{Lewis:1999bs}, a widely used Boltzmann solver in cosmology. Our implementation is designed to maintain consistency and compatibility with the latest version of \texttt{CAMB}, with a focus on numerical stability and computational efficiency. We also demonstrate how this new implementation can be integrated into Markov Chain Monte Carlo (MCMC) frameworks to obtain updated cosmological constraints on ULAs from CMB measurements.  Figure \ref{fig:m_axion_vs_f_axion} shows our main result, the marginalised distribution of the axion density, $\Omega_{\mathrm{ax}}\,h^2$, for fixed values of the axion mass in the range $10^{-28}-10^{-24}$ eV  using our implementation of the Passaglia--Hu 
EFA combined with \Planck\ PR4 data~\cite{rosenberg22} and  DESI BAO~\cite{DESI:2024mwx}.

The remainder of this paper is organized as follows. In Sec.~\ref{sec:EFA}, we summarize the standard and Passaglia-Hu EFA, emphasizing how auxiliary fields facilitate the removal of fast oscillations. Sec.~\ref{sec:camb} provides a detailed description of our numerical implementation within \texttt{CAMB}, including discussions on numerical stability, validation tests, boundary conditions, and the transition between the auxiliary-field description and the effective fluid representation. Sec.~\ref{sec:constraints} outlines the integration of our implementation with the \texttt{Cobaya} MCMC framework and presents preliminary constraints derived from CMB data. Finally, we conclude with a discussion of potential extensions and future directions for this work.

\section{Overview of the Effective Fluid Approximation}
\label{sec:EFA}

\subsection{Klein-Gordon Evolution}
\label{sec:KG}

The full evolution of the axion field, $\phi(\vec{x}, t)$, is governed by the Klein-Gordon (KG) equation:
\begin{equation}
    \Box \phi(\vec{x}, t) = \frac{{\rm d} V(\phi(\vec{x}, t))}{{\rm d} \phi} 
    \approx m^2 \phi(\vec{x}, t)\,,
\end{equation}
where $m$ is the axion mass and we are making the usual assumption that the field is sufficiently close to the minimum that the axion cosine potential can be approximated by $V(\phi) \approx \frac{1}{2} m^2 \phi^2$.

In a homogeneous and isotropic Friedmann–Lemaître–Robertson–Walker (FLRW) cosmology, the background equation of motion becomes
\begin{equation}
    \ddot\phi + 2 
    {\cal H} 
    \dot\phi + a^2 m^2 \phi = 0\,,
\end{equation}
where overdots denote derivatives with respect to conformal time $\eta = \int {\rm d}t/a$, $a$ is the scale factor, and ${\cal H} = a H$ is the conformal Hubble parameter.

The energy density, $\rho_{\text{ax}}$, and pressure, $P_{\text{ax}}$, of the axion field are given by
\begin{align}
    \rho_{\text{ax}} & =  \frac{1}{2} \frac{\dot{\phi}^2}{a^2} + \frac{1}{2} m^2 \phi^2 \,, \\
    P_{\text{ax}} & = \frac{1}{2} \frac{\dot{\phi}^2}{a^2} - \frac{1}{2} m^2 \phi^2 \,.
\end{align}

The perturbed Klein-Gordon (KG) equation in the synchronous gauge is given by
\begin{equation}
    \ddot{\delta\phi} + 2 {\cal H} \dot{\delta\phi} + (k^2+a^2 m^2) \delta\phi = - \frac{\dot h}{2} \dot\phi\,,
\end{equation}
where $k$ is the co-moving wavenumber and $h$ is the trace of the spatial metric perturbation. The axion density and pressure perturbations are defined as
\begin{align}
    \delta \rho_{\text{ax}} &= a^{-2} \dot\phi \dot{\delta\phi} + m^2 \phi \delta\phi\,,\\
    \delta P_{\text{ax}} & = a^{-2} \dot\phi \dot{\delta\phi} - m^2 \phi \delta\phi\,, \\
    \rho_{\text{ax}} u_{\text{ax}} &= a^{-2} k \dot\phi \delta\phi\,,
\end{align}
where $u_\text{ax} \equiv (1+w_\text{ax}) v_\text{ax}$ is the axion heat-flux, and $w_\text{ax} = P_\text{ax} / \rho_\text{ax}$ is the equation of state.

\subsection{Standard EFA}
\label{sec:standard-EFA}

The standard EFA, used in codes like \texttt{AxionCAMB}~\cite{Hlozek:2014lca,Hlozek:2017zzf} and \texttt{AxiCLASS}~\cite{Poulin:2018dzj}, involves evolving the full KG equation until the field starts oscillating ($m/H \gtrsim 1$). At $a=a_\star$, defined by $m \approx 3\,H(a_\star)$, a transition is made from solving the full KG equation to a fluid description. Specifically, fluid variables $(\rho_{\text{ax}},\,\delta_{\text{ax}},\,u_{\text{ax}})$ are initialized using values derived from the KG quantities at $a_\star$, with $\delta_{\text{ax}}(a_\star)= \delta\rho_{\text{ax}}(a_\star) / \rho_{\text{ax}}(a_\star)$. The factor of 3 in the definition of $a_\star$ is a numerical choice that balances accuracy and computational efficiency.

After initialization, the evolution of the density contrast, $\delta_{\text{ax}}$, and heat flux, $u_{\text{ax}}$, for a given mode $k$ is governed by the standard fluid equations in the synchronous gauge
\begin{eqnarray} \label{eqn:fluid_perts}
    \dot{\delta}_\text{ax}&=&-\bigg[k u_\text{ax}+(1+w_\text{ax})\frac{\dot{h}}{2}\bigg] \nonumber\\ & & -3 {\cal H} (c_\text{ax}^2-w_\text{ax})\left( \delta_\text{ax} + 3 {\cal H} \frac{u_\text{ax}}{k} \right) \nonumber\\
     & & - 3 {\cal H} \frac{ \dot{w}_\text{ax}}{(1+w_\text{ax}) } \frac{ u_\text{ax}}{k}  \,,\\
    \dot{u}_\text{ax}&=&-(1-3c_\text{ax}^2){\cal H}u_\text{ax}+  \frac{\dot{w}_\text{ax}}{(1+w_\text{ax}) } u_\text{ax} \nonumber\\ && + k c_\text{ax}^2\, \delta_\text{ax}\,,
\end{eqnarray}
where $c_\text{ax}$ is the sound speed in the axion rest frame.

In the \emph{standard} EFA, the axion is treated as a pressureless fluid ($w_\text{ax}=\dot{w}_\text{ax}=0$), but with a scale-dependent sound speed:
\begin{equation}
c_\text{ax}^2 = \frac{k^2/(4 m^2 a^2)}{1 + k^2/(4 m^2 a^2)}\,.
\end{equation}
At late times, this captures the average effect of the rapid axion field oscillations without requiring direct solution of the oscillatory KG equation. The approximation works reasonably well because once $m \gg H$, the residual pressure support of the field is effectively described by this $(k,a)$-dependent sound speed, and the time-averaged energy density behaves like non-relativistic matter.

\subsection{Passaglia and Hu EFA}
\label{sec:passaglia-hu}

Passaglia and Hu~\cite{Passaglia:2022bcr} introduced an improved effective fluid approximation, which we refer to as the PH EFA. To facilitate numerical computation, they decompose the axion field into two auxiliary fields, $\varphi_c$ and $\varphi_s$:
\begin{equation} \label{eqn:background_decomp}
    \phi(\tau) = \varphi_c(\tau) \cos\left[\tau -\tau_* \right] + \varphi_s(\tau) \sin\left[\tau - \tau_* \right]\,,
\end{equation}
where $\tau=mt$ is a dimensionless time variable and $\tau_*$ is a reference time (the switch time) in the oscillatory regime.

These auxiliary fields represent the slowly varying envelope of the axion field, effectively filtering out the rapid oscillations. Substituting Eq.~\eqref{eqn:background_decomp} into the KG equation yields the equations of motion for $\varphi_c$ and $\varphi_s$ gives
\begin{subequations}
\begin{align}
    \varphi_c'' + 2 \varphi_s' + 3 \frac{H}{m} [\varphi_s + \varphi_c' ]  \ &=0\,,\\  \varphi_s''- 2 \varphi_c' + 3 \frac{H}{m} [-\varphi_c + \varphi_s']  &= 0\,,
\end{align}
\end{subequations}
where primes denote derivatives with respect to $\tau$. 

The \emph{effective} fluid energy density, $\rho_{\text{ax}}^{\text{ef}}$, and pressure, $P_{\text{ax}}^{\text{ef}}$, are defined as
\begin{align}
    \label{eqn:effective_rho}
    \rho_{\text{ax}}^{\text{ef}} &\equiv  \frac{1}{2} m^2 \left(\varphi_c^2 +  \varphi_s^2  + \frac{{\varphi_c'}^2}{2} +\frac{{\varphi_s'}^2}{2}  -  \varphi_c \varphi_s' +  \varphi_s \varphi_c'\right)\,,  \\
    P_{\text{ax}}^{\text{ef}} &\equiv \frac{1}{2} m^2 \left(\frac{{\varphi_c'}^2}{2} +\frac{{\varphi_s'}^2}{2}  -  \varphi_c \varphi_s' +  \varphi_s \varphi_c'\right)\,,
\end{align}
which satisfy the usual fluid conservation law. Analysis of the ratio $P_{\text{ax}}^{\text{ef}}/\rho_{\text{ax}}^{\text{ef}}$ in the late-time oscillatory regime reveals an approximate equation of state
\begin{equation}
w_{\text{ax}} = \frac{3}{2} \left(\frac{m}{H}\right)^{-2}\,.
\end{equation}

To minimize matching errors and suppress fast oscillatory modes at the switch time, specific matching conditions are imposed. These constrain the ratio of the second and first derivatives of the auxiliary fields $\varphi_{c,s}$ (collectively denoted) at the switch time: $\left.\frac{\varphi_{c,s}''}{\varphi_{c,s}'}\right\vert_{*}=D $, where
\begin{equation}
\label{eq:constraints}
D \equiv \left.-\frac{1}{2} \frac{\langle H \rangle}{m} \left(3 - \frac{m}{\langle H\rangle ^3}\frac{d\langle H\rangle^2}{d \tau}\right)\right\vert_{*} \,,
\end{equation}
and $\langle H \rangle$ represents the Hubble rate averaged over axion oscillation cycles. In practice, instead of averaging, an effective Hubble rate $H^{\text{ef}}$, sourced by $\rho_{\text{ax}}^{\text{ef}}$, is used.

To initialize the effective fluid approximation, the auxiliary field variables $\varphi_{c*}$, $\varphi_{s*}$, $\varphi_{c*}'$, and $\varphi_{s*}'$ at the switch time are determined by solving the matching conditions

\begin{subequations}
\label{eqn:background_aux}
\begin{align}
\varphi_{c*}
&=
\phi_{*},
\\[10pt]
\varphi_{s*}
&=
\dfrac{
  \phi_{*}'\,\bigl[\bigl(D\,m + 3\,H_*\bigr)^2 + 4\,m^2\bigr]
  + 6\,H_*\,m\,\phi_{*}
}{
  D^2\,m^2 + 3\,D\,H_*\,m + 4\,m^2
},
\\[10pt]
\varphi_{s*}'
&=
\dfrac{
  3\,H_*\,\bigl[-2\,\phi_{*}' + D\,\phi_{*}\bigr]
}{
  D^2\,m + 3\,D\,H_* + 4\,m
},
\\[10pt]
\varphi_{c*}'
&=
-\,\dfrac{
  3\,H_*\,\Bigl[D\,m\,\phi_{*}' + 3\,H_*\,\phi_{*}' + 2\,m\,\phi_{*}\Bigr]
}{
  D^2\,m^2 + 3\,D\,H_*\,m + 4\,m^2
}\,,
\end{align}
\end{subequations}
where starred quantities are evaluated at $\tau_*$.

Similarly, the axion perturbation $\delta \phi$ is decomposed into two parts, $\delta \varphi_c$ and $\delta \varphi_s$
\begin{equation}
  \delta \phi(\tau) = \delta \varphi_c(\tau) \cos[\tau - \tau_*] + \delta \varphi_s(\tau) \sin[\tau - \tau_*].
\label{25}
\end{equation}
The equations of motion for $\delta\varphi_c$ and $\delta\varphi_s$ are
\begin{subequations}
\begin{align}
\delta \varphi_c'' 
&\;+\; 2 \delta \varphi_s'
\;+\; 3 \frac{H}{m}\bigl( \delta \varphi_c' + \delta \varphi_s \bigr)
\;+\; \frac{k^2}{a^2 m^2}\,\delta \varphi_c 
\nonumber\\
&=\; -\tfrac{h'}{2}\,\bigl(\varphi_c' + \varphi_s \bigr)\,,
\\[6pt]
\delta \varphi_s'' 
&\;-\; 2 \delta \varphi_c'
\;+\; 3 \frac{H}{m}\bigl( -\,\delta \varphi_c + \delta \varphi_s' \bigr)
\;+\; \frac{k^2}{a^2 m^2}\,\delta \varphi_s 
\nonumber\\
&=\; -\tfrac{h'}{2}\,\bigl(\varphi_s' - \varphi_c \bigr)\,.
\end{align}
\end{subequations}

The effective fluid versions of the perturbed quantities are defined as
\begin{align}
    \delta \rho_{\text{ax}}^{\text{ef}}   ={}&\ \frac{1}{2} m^2 \left[\varphi_s \delta\varphi_c' - \varphi_c \delta\varphi_s' + \delta\varphi_c' \varphi_c'+ \delta\varphi_s' \varphi_s' \right.\\&+ \left.  \delta\varphi_s (2 \varphi_s + \varphi_c') + \delta\varphi_c  (2 \varphi_c - \varphi_s') \right]\,,\\
    \delta P_{\text{ax}}^{\text{ef}}  ={}&\ \delta\rho_{\text{ax}}^{\text{ef}} - m^2 \left[\delta \varphi_s \varphi_s + \delta \varphi_c \varphi_c \right]\,,\\
    \rho_{\text{ax}}^{\text{ef}}  u_{\text{ax}}^{\text{ef}} ={}&\ \frac{k m}{2 a} \left[\delta \varphi_c \left(\varphi_s+\varphi_c'\right)+ \delta \varphi_s\left(-\varphi_c+\varphi_s'\right)\right]\,,
\end{align}

Similar to the background, the perturbation matching condition $\left.\frac{\delta \varphi_{c,s}''}{\delta \varphi_{c,s}'}\right\vert_{*}=D $ is used to suppress oscillatory modes. Solving for the auxiliary variables at the switch provides the initial conditions for the fluid perturbations 

\begin{subequations}
\label{eqn:perturb_aux}
\begin{widetext}
\begin{align}
\delta\varphi_{c*} 
&= \delta\phi_{*} \,,
\\[1em]
%
\delta\varphi_{s*}
&=
\frac{
  \begin{aligned}
    & \Bigl( 2\,\delta\phi_* 
      \Bigl[
        D\,k^2\,m 
        + 3\,H_*\bigl(k^2 + 2\,a^2\,m^2\bigr)
      \Bigr]
    \\
    &\quad
    +\,a^2\,m 
      \Bigl\{
        2\,\delta\phi' \Bigl[
          9\,H_*^2 
          + 6\,D\,H_*\,m 
          + \bigl(4 + D^2\bigr)\,m^2
        \Bigr]
        \\
        &\qquad
        +\,h_s'\,m 
          \Bigl(
            3\,\varphi_{c*}'\,H_*
            + D\,\varphi_{c*}'\,m
            - 2\,\varphi_{s*}'\,m
            + 2\,m\,\varphi_{c*}
            + 3\,H_*\,\varphi_{s*}
            + D\,m\,\varphi_{s*}
          \Bigr)
      \Bigr\} \Bigr)
  \end{aligned}
}{
  \begin{aligned}
    2\,m
    \Bigl[
      2\,k^2
      + a^2\,m \Bigl(
        3\,D\,H_*
        + 4\,m
        + D^2\,m
      \Bigr)
    \Bigr]
  \end{aligned}  
} \,,
\\[1em]
%
\varphi_{s*}' 
&=
\frac{
  \begin{aligned}
    & \Bigl(-2\,\delta\phi_*
      \Bigl(
        k^4 
        + 2\,a^2\,k^2\,m^2
        - 3\,a^4\,D\,H_*\,m^3
      \Bigr)
    \\
    &\quad
    -\,a^2\,m 
      \Bigl\{
        2\,\delta\phi_*'
         \Bigl[
           D\,k^2\,m
           + 3\,H_*\bigl(k^2 + 2\,a^2\,m^2\bigr)
         \Bigr]
        \\
        &\qquad
        +\,h_s'\,m
         \Bigl[
           \varphi_{c*}'\bigl(k^2 + 2\,a^2\,m^2\bigr)
           + k^2\,\varphi_{s*}
           + a^2\,m^2\bigl(
              D\,\varphi_{s*}'
              - D\,\varphi_{c*}
              + 2\,\varphi_{s*}
            \bigr)
         \Bigr]
      \Bigr\} \Bigr)
  \end{aligned}
}{
  \begin{aligned}
    2 a^2\,m^2\Bigl[
      2\,k^2
      + a^2\,m \bigl(
        3\,D\,H_*
        + 4\,m
        + D^2\,m
      \bigr)
    \Bigr]
  \end{aligned}
}  \,,
\\[1em]
%
\varphi_{c*}' 
&=
\frac{
  \begin{aligned}
    & \Bigl\{ 4\,\delta\phi_*'\,k^2\,m 
    \\
    &\quad
     -\,2\,\delta\phi_*
      \Bigl[
        D\,k^2\,m 
        + 3\,H_*\bigl(k^2 + 2\,a^2\,m^2\bigr)
      \Bigr]
    \\
    &\quad
     -\,a^2\,m 
      \Bigl[
        6\,\delta\phi_*'\,H_*\bigl(3\,H_* + D\,m\bigr)
        \\
        &\qquad
        +\,h_s'\,m 
         \Bigl(
           3\,\varphi_{c*}'\,H_* 
           + D\,\varphi_{c*}'\,m 
           - 2\,\varphi_{s*}'\,m
           + 2\,m\,\varphi_{c*}
           + 3\,H_*\,\varphi_{s*}
           + D\,m\,\varphi_{s*}
         \Bigr)
      \Bigr] \Bigr\}
  \end{aligned}
}{
  \begin{aligned}
    2 m \Bigl[
      2\,k^2
      + a^2\,m
      \bigl(
        3\,D\,H_*
        + 4\,m
        + D^2\,m
      \bigr)
    \Bigr]
  \end{aligned} 
}  \,.
\end{align}
\end{widetext}
\end{subequations}

At the switch, the fluid variables $(\rho_{\text{ax}},\,\delta_{\text{ax}},\,u_{\text{ax}})$ are initialized using the effective values derived from the auxiliary variables.

The EFA is completed by specifying the sound speed for perturbations
\begin{equation}
    c_{ax}^2 = \left(\frac{k}{a m}\right)^{-1} \left( \sqrt{1+\left(\frac{k}{a m} \right)^2} -1 \right) + \frac{5}{4} \left(\frac{m}{H}\right)^{-2}\,,
\end{equation}
where the final term again calibrates the true KG solution.

\section{CAMB Implementation} \label{sec:camb}

\subsection{Background Evolution}

Our implementation begins by evolving the homogeneous axion field, $\phi(a)$, using $\ln a$ as the integration variable. The second-order KG equation is rewritten as two coupled first-order equations
\begin{align}
  \frac{{\rm d}\phi}{{\rm d}\ln a} &= \frac{\dot{\phi}}{\mathcal{H}}, \\
  \frac{{\rm d}}{{\rm d}\ln a} \bigl[a^2 \dot{\phi}\bigr] &= -\,\frac{a^2}{\mathcal{H}}\,\frac{{\rm d} V}{{\rm d}\phi},
\end{align}
where $V(\phi)=\tfrac12\,m^2 \phi^2$. These equations are numerically integrated from early times ($a\ll1$) with a specified initial field value, $\phi_\text{i}$, until the axion field enters the oscillatory regime. The switch to the EFA can be triggered by several criteria: when $m/H_* > \beta$ (with $\beta$ being a user-defined parameter), after a specified number of oscillation cycles, or at a predefined scale factor $a_*$. After the switch, the code continues to evolve $\phi$ using the KG equation for a short ``grace period" (approximately one e-fold in $\ln a$) to ensure a smooth transition to the fluid description.

At the transition to the EFA, the auxiliary variables are initialized using Eqs.~\ref{eqn:background_aux}. Subsequently, the effective fluid density, which we denote by $\rho_{\mathrm{ax,\,fluid}}$, is initialized according to Eq.~\ref{eqn:effective_rho}. The computation of the auxiliary variables involves solving an implicit system of equations, which we perform iteratively, using the values from KG solution as an initial guess. This process converges within a few iterations.

Following the switch, the effective fluid density, $\rho_{\mathrm{ax,\,fluid}}$, is evolved using the standard conservation equation:
\begin{equation} 
  \frac{{\rm d}\rho_{\mathrm{ax,\,fluid}}}{{\rm d}\ln a}
  =
  -\,3\bigl[1 + w_{\mathrm{ax}}(a)\bigr]\,\rho_{\mathrm{ax,\,fluid}}(a),
  \label{eq:axion_fluid_conservation}
\end{equation}
where the equation of state $w_{\mathrm{ax}}(a)$ can be chosen to match either the standard EFA or the Passaglia-Hu prescription, yielding a slowly varying $w_{\mathrm{ax}} \sim (m/H)^{-2}$.

To ensure a smoother transition, we combine the exact scalar-field density and the effective fluid density using a weighting function
\begin{equation} \label{eqn:weighting}
  \rho_{\mathrm{ax}}(a) =
    W(a)\,\rho_{\mathrm{ax,\,KG}}(a)\,+\,\bigl[\,1 - W(a)\bigr]\,\rho_{\mathrm{ax,\,fluid}}(a),
\end{equation}
where $W(a)= \exp\bigl[\alpha\,(\ln a_* - \ln a)\bigr]$ and $\alpha>0$ controls the sharpness of the transition. For $a\ll a_*$, the evolution is dominated by the exact field solution, while for $a\gg a_*$, the fluid limit is recovered. The field variables $\{\phi(a),\dot{\phi}(a)\}$, and $\rho_{\mathrm{ax}}(a)$ at the switch point, $a_{\star}$, are stored and used for spline interpolation when evolving perturbations.

A practical aspect of implementing ULA models is the common requirement to fix the present-day axion density, $\Omega_{\mathrm{ax}}\,h^{2}$. Our implementation addresses this by iteratively searching for the initial field value $\phi_{i}$ (at a very early time $a_i \ll 1$) that yields the desired $\Omega_{\mathrm{ax}}\,h^{2}$. This involves a root-finding procedure in $\phi_{i}$, where the background field equation is repeatedly integrated from $a_{\mathrm{start}}$ up to $a=1$ until the final axion density matches the target. Since this step only involves the background equation, the computational cost is minimal. 

\begin{figure}
    \includegraphics[width=\columnwidth]{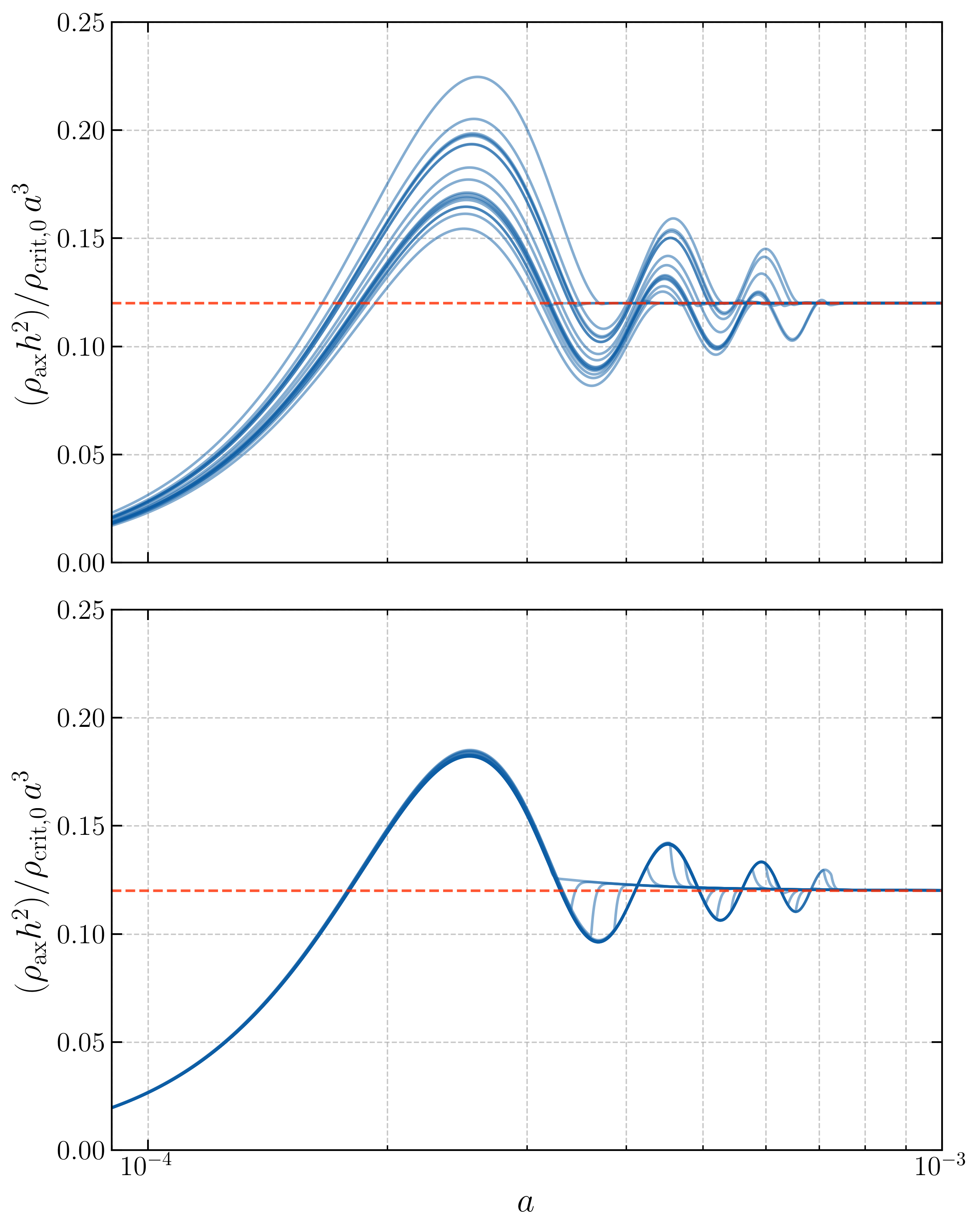}
    \caption{Evolution of the axion energy density around the time it starts oscillating, for an axion mass of $m = 10^{-27}\,\mathrm{eV}$ and $\Omega_{\mathrm{ax}}\,h^{2}=0.12$. The top panel shows the evolution using the standard Effective Fluid Approximation (EFA) for different switch points, defined by $m/H_*$ ranging from 5 to 20. The bottom panel displays the evolution using the Passaglia-Hu EFA. In both panels, the vertical axis shows the axion energy density scaled by the present-day critical density and $a^3$, and the horizontal axis is the scale factor $a$. The dashed horizontal line indicates the target energy density. The weighting factor for the transition between the Klein-Gordon and fluid descriptions, Eq.~(\ref{eqn:weighting}), is $\alpha = 100$.}
    \label{fig:background_phi}
\end{figure}

As a check of the background dynamics for the standard EFA versus the PH EFA, we set $\Omega_{\mathrm{ax}}\,h^{2}=0.12$. This is illustrated in Fig.~\ref{fig:background_phi} for an axion mass of $10^{-27}\,\mathrm{eV}$. The top panel of Fig.~\ref{fig:background_phi} shows the evolution for the standard EFA with the switch point, defined by $m/H_*$, ranging from 5 to 20. The weighting factor for the transition between the KG and fluid descriptions is set to $\alpha = 100$. For the standard EFA, achieving the target present-day density requires a significant variation in the initial field value, with a difference of approximately 3.9\%. This sensitivity to the choice of switch point results in substantial changes to the axion density near matter-radiation equality. 

In contrast, the bottom panel of Fig.~\ref{fig:background_phi} demonstrates the robustness of the PH EFA, requiring a much smaller variation in the initial field value, only around 0.2\%, to reach the same target density across the same range of switch points. This improved stability is attributed to the auxiliary fields removing oscillations, and the higher accuracy of the equation of state, $w_{\mathrm{ax}}$, which is accurate to order $(m/H)^{-2}$. Extending the range of $m/H_*$ for the switch point from 10 to 20 results in an even smaller variation in the initial field value, of only about 0.05\%.

\subsection{Perturbations}

\begin{figure*}
    \includegraphics[width=\textwidth]{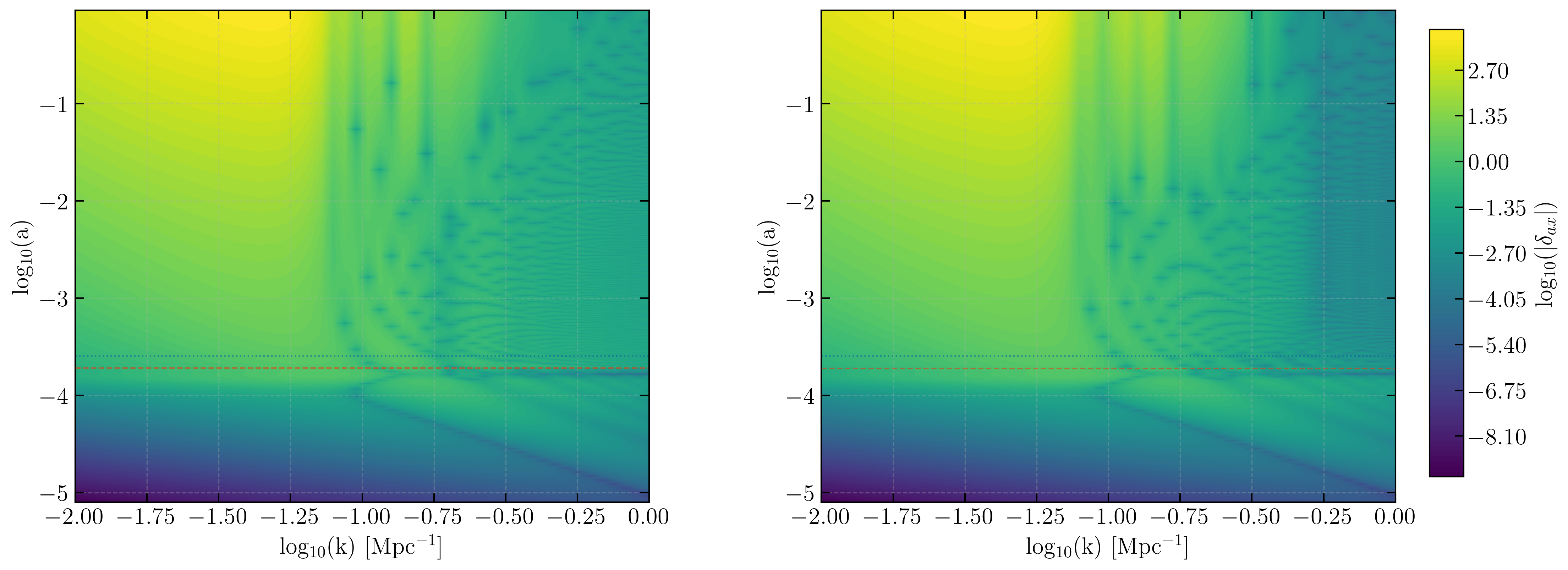}
    \caption{The axion density, $\delta_\text{ax}$, as a function of scale factor, $a$, and wavenumber, $k$, for an axion mass of $m = 5 \times 10^{-27}\,\mathrm{eV}$ and a switch at $m/H_*=10$. The switch time is indicated by the dashed red line, and matter-radiation equality by the dotted blue line. The PH EFA yields a more accurate evolution of $\delta_\text{ax}$, even when switching relatively early in the oscillatory regime. \textbf{Left:} Standard EFA. \textbf{Right:} PH EFA. }
    \label{fig:contour_plot_comparison_5e-27_10}
\end{figure*}

\begin{figure*}
    \includegraphics[width=\textwidth]{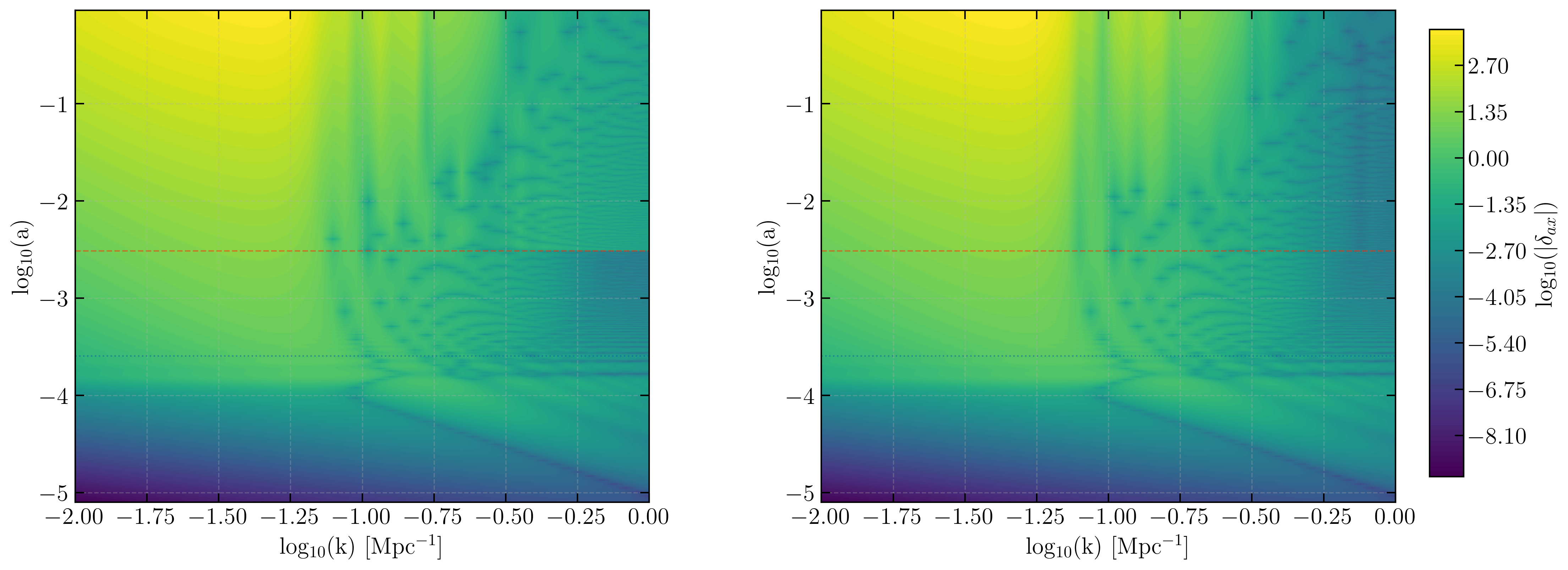}
    \caption{As for Fig.~\ref{fig:contour_plot_comparison_5e-27_10} but for a switch $m/H_*=1000$.}
    \label{fig:contour_plot_comparison_5e-27_1000}
\end{figure*}

The implementation of perturbations closely follows the background evolution strategy. We evaluate the auxiliary fields $\delta\varphi_c$ and $\delta\varphi_s$ at the switch time, at which point we transition to solving the effective fluid equations. The fluid perturbations, $\delta_{\text{ax}}$ and $u_{\text{ax}}$, are initialized at $a_*$ using the matching conditions derived from the auxiliary field values as described in Sec.~\ref{sec:passaglia-hu}. Following the switch, we continue to evolve the KG equation for a short grace period and apply the same  weighting as in Eq.~\ref{eqn:weighting} to smoothly transition between the auxiliary field solution and the effective fluid description. This weighting is applied to the density and heat flux perturbations, $(\,\delta_{\text{ax}},\,u_{\text{ax}})$.

Figures~\ref{fig:contour_plot_comparison_5e-27_10} and~\ref{fig:contour_plot_comparison_5e-27_1000} compare the evolution of the axion density contrast, $\delta_{\text{ax}}$, for the standard EFA (left panels) and the PH EFA (right panels) as a function of scale factor, $a$, and comoving wavenumber, $k$. We show results for an axion mass $m=5\times10^{-27} \, \mathrm{eV}$, with two choices of the switch point: a relatively early transition at $m/H_{*}=10$ (Fig.~\ref{fig:contour_plot_comparison_5e-27_10}) and a later transition at $m/H_{*}=1000$ (Fig.~\ref{fig:contour_plot_comparison_5e-27_1000}). The $m/H_{*}=1000$ case can be viewed as a ground-truth benchmark since the field is deep in the oscillatory regime at the time of switching. This particular mass is chosen as matter-radiation equality is close to the $m/H_{*}=10$ switch point.

On large-scales, ULAs behave like a non-relativistic component after oscillations begin; at small scales, however, there is a scale-dependent pressure support introduced by the field’s oscillatory nature. In the standard EFA case (left), one observes that immediately after the switch time (marked by the red dashed line), the evolution of $\delta_\text{ax}$
  can develop spurious oscillatory features or mismatches in amplitude. The mismatch in amplitude is apparent on small-scales even for $m/H_{*}=1000$, demonstrating that late switches can still be problematic. 
  
  By contrast, the PH EFA case (right) highlights a smoother and more robust evolution of $\delta_\text{ax}$, with the fluid approximation much closer to the the full KG evolution. This reduces the need for fine-tuning the switch time, which is often a nuisance parameter in the standard EFA. An important consequence of these differences emerges in observable signatures. For axion masses $\sim 10^{-27} - 10^{-28}\,\mathrm{eV}$, the field commences oscillating around matter-radiation equality. Small inaccuracies in the predicted timing of these oscillations, or in the residual pressure support, can shift how structures grow on sub-horizon scales. 

\begin{figure*}
    \includegraphics[width=\textwidth]{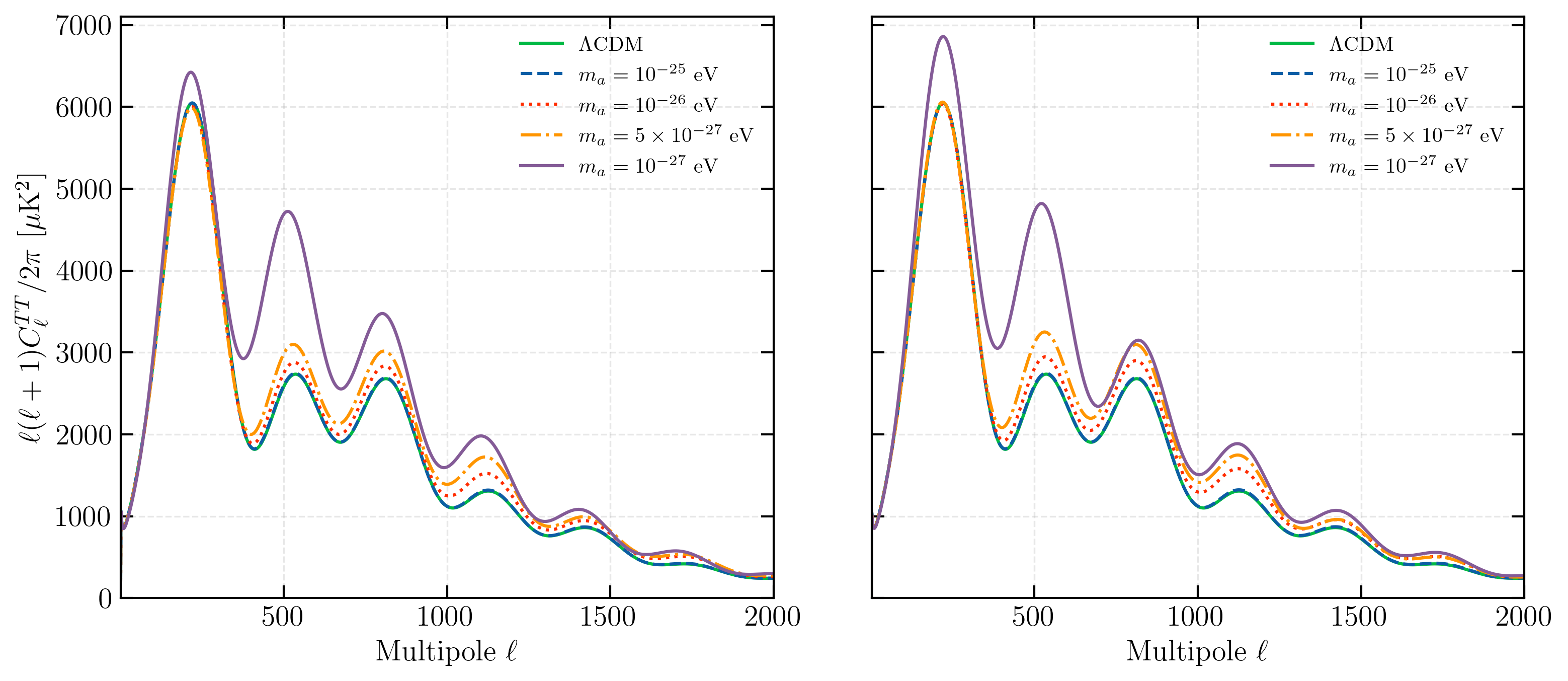}
    \caption{The CMB temperature power spectra ($C_{\ell}^{TT}$) for a switch $m/H_*=3$, comparing \lcdm\ (black) and axion models with masses $m \in \{5\times 10^{-27},\, 10^{-26},\, 5\times10^{-26},\,10^{-25}\}\,\mathrm{eV}$. The fraction of dark matter in axions is set to $f_\text{ax} = 1$, i.e.\ all of the dark matter is composed of axions. \textbf{Left:} Standard EFA. \textbf{Right:} PH EFA. }
    \label{fig:cls_3}
\end{figure*}

\begin{figure*}
    \includegraphics[width=\textwidth]{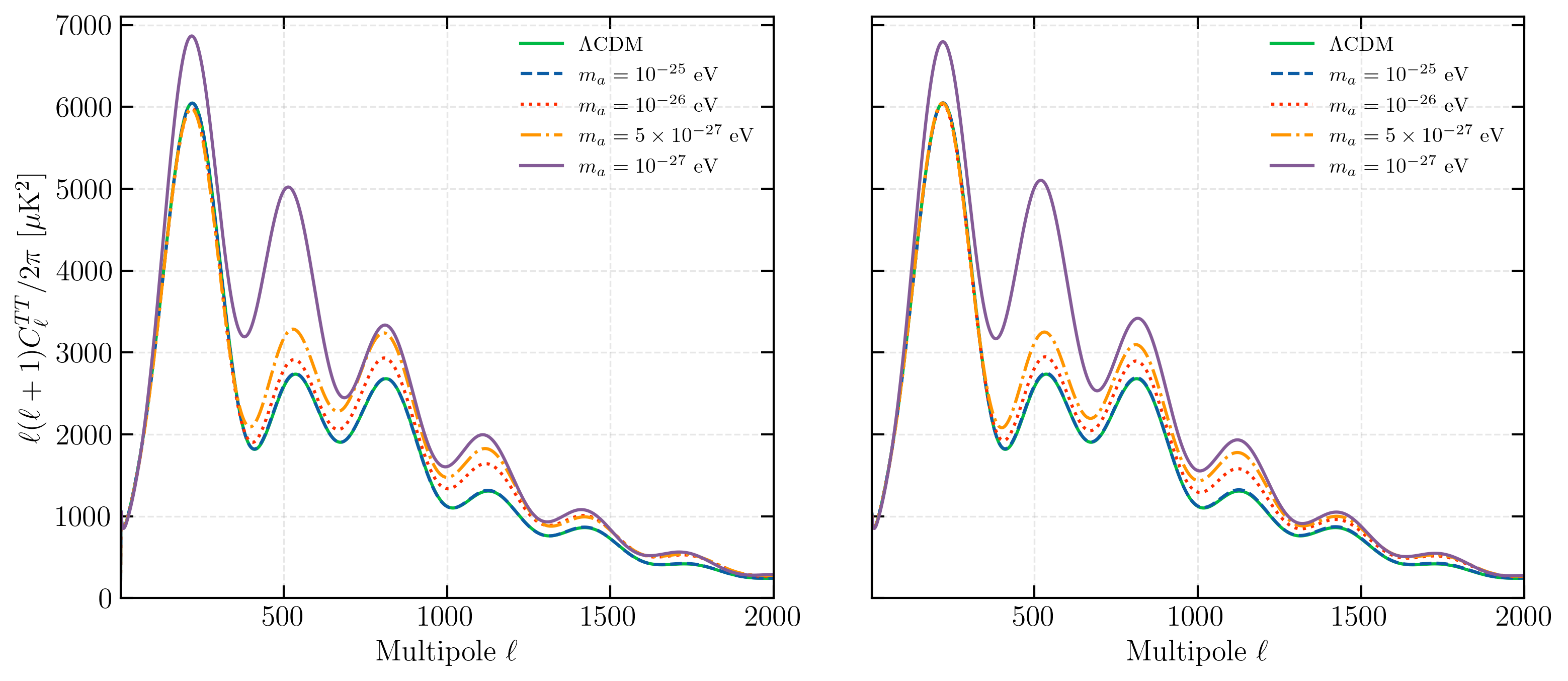}
    \caption{As for Fig.~\ref{fig:cls_10} but for a switch $m/H_*=10$.}
    \label{fig:cls_10}
\end{figure*}

The Cosmic Microwave Background (CMB) power spectrum is a key probe of cosmology, sensitive to the effects of dark matter near recombination. We therefore use this as a primary benchmark to compare the accuracy and performance of the standard and improved PH EFA. Figures~\ref{fig:cls_3} and \ref{fig:cls_10} show the CMB temperature power spectrum ($C_{\ell}^{TT}$) for models with switch values of $m/H_*=3$ and $m/H_*=10$ respectively. The left panels in both figures show the power spectrum calculated using the standard EFA, while the right panels show results from the PH EFA. In each panel we show $\Lambda$CDM, along with axion  masses ranging from  $m = 10^{-27}$ eV to $10^{-25}\,\mathrm{eV}$. We define the fraction of dark matter composed of axions, $f_\text{ax}$, as $f_\text{ax} = \Omega_\text{ax} / \Omega_\text{D}$, where $\Omega_\text{D}= \Omega_\text{ax} + \Omega_\text{CDM}$ is the total dark matter density. When comparing with $\Lambda$CDM, we set $f_\text{ax}=1$, and otherwise use \Planck\ 2018 best-fit parameters~\cite{Planck:2018vyg}.

The PH EFA better captures the small-scale damping signature arising from ULA oscillations. In particular, for axion masses near $10^{-27}\,\mathrm{eV}$, the standard EFA can misestimate the suppression of temperature power at high $\ell$ when switching from the KG to fluid description too early. The PH EFA requires a  less aggressive (later) switch to maintain comparable accuracy, reducing potential parameter biases when fitting to data. Even at $m/H_*=100$, there are visible changes for the standard EFA.

Since the presence of ULAs modifies the expansion history, this impacts the reonization history. The change in $H(z)$ is taken into account in \texttt{recfast}~\cite{Seager:1999bc}. It also changes the evolution of the matter-temperature, $T_\text{M}$, through ${\rm d} H/{\rm d} z$~\cite{Scott:2009sz}, which is not automatically taken into account in \texttt{recfast}.  However, for the range of axion masses and abundances considered here, the induced changes in ${\rm d} H/{\rm d} z$ are small enough that they do not significantly alter the visibility function relative to the standard $\Lambda$CDM case. Empirically, we verified that modifying \texttt{recfast} to account for these variations has only a negligible effect on the CMB anisotropies at the accuracy levels we are concerned with. Hence, for practical purposes, we do not include such corrections in our code.

\subsection{Optimising the Switch}

The optimal switch should balance computational speed and accuracy. For large values of $m / H_*$, higher accuracy settings in \texttt{CAMB} are required to maintain the fidelity of the KG solution. Empirically, we find that when $m / H_* \gtrsim 100$, the default setting (\texttt{accuracy=1}) for the Runge-Kutta integrator can be insufficient.

We define a normalised compute time, $T = 1$, using parameters $m = 10^{-27}\,\mathrm{eV}$, $m / H_* = 100$, $f_\text{ax} = 1$, \texttt{accuracy=1}, and other \Planck\ 2018 best-fit values. Increasing \texttt{accuracy} to 2 and 3 raises $T$ to approximately 1.3 and 1.7, respectively. Doubling or tripling $m / H_*$ to 200 and 300 with \texttt{accuracy=1} leads to only a modest increase in compute time, $T \approx 1.05$ and 1.1. However, the perturbed KG equation can diverge at large $m / H_*$, especially for small $k$ modes, necessitating higher accuracy settings and longer runtimes. This makes it preferable to keep $m / H_*$ as small as possible while using default accuracy settings.

As a ground-truth reference, we adopt the PH EFA with a switch at $m / H_* = 300$ and \texttt{accuracy=3}. For earlier switch times and axion masses in the range $10^{-28}\,\mathrm{eV}$ to $10^{-25}\,\mathrm{eV}$, we compute the resulting CMB temperature power spectra, $C_{\ell}^{(\text{model})}$, using \texttt{accuracy=1}. These are compared to the ground-truth spectra, $C_{\ell}^{(\text{ref})}$, via the mismatch measure:
\begin{eqnarray}
\chi^2 
&=& 
\sum_{\ell=2}^{\ell_{\mathrm{max}}}
\left(\frac{C_{\ell}^{(\text{model})} - C_{\ell}^{(\text{ref})}}{\sigma_{C_{\ell}}}\right)^2 \,, \\ \nonumber
\quad
\sigma_{C_{\ell}} &=&
\sqrt{\frac{2}{2\ell + 1}}\;C_{\ell}^{(\text{ref})} \,,
\end{eqnarray}
assuming a cosmic-variance-limited experiment with $\ell_{\mathrm{max}} = 2000$.

\begin{figure*}
    \includegraphics[width=\textwidth]{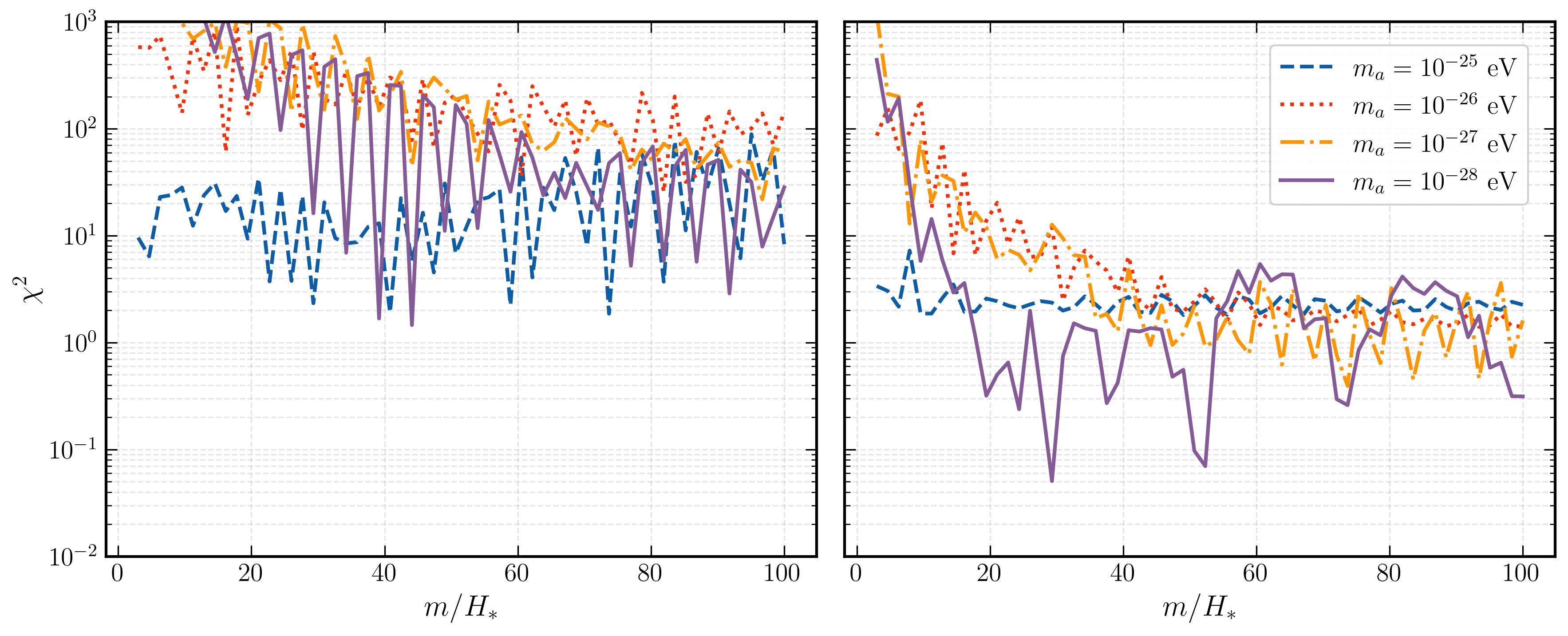}
    \caption{Comparison of the mismatch measure $\chi^2$ between model and ground-truth CMB temperature power spectra as a function of $m/H_*$ for axion masses $m=10^{-25},\,10^{-26},\,10^{-27},\,10^{-28}\,\mathrm{eV}$. \textbf{Left:} Standard EFA. \textbf{Right:} PH EFA.}
    \label{fig:chi2}
\end{figure*}

Figure~\ref{fig:chi2} illustrates $\chi^2$ as a function of $m/H_*$ for four representative axion masses. The left panel corresponds to the standard EFA, while the right panel shows the PH EFA. Larger inaccuracies arise for masses where the field begins oscillating near matter-radiation equality. The non-smooth behavior stems from errors depending on the phase of the oscillation cycle at the switch point.

In the standard EFA, $\chi^2$ values remain high unless $m / H_*$ is very large, indicating significant biases when switching too early during the oscillatory phase. Conversely, in the PH EFA, $\chi^2$ is significantly lower across a broader range of switch times. With $m / H_* \approx 40$–$50$, the PH EFA maintains close agreement with the late-switch baseline, keeping $\chi^2$ near unity in most cases. This robustness to earlier switches results from the auxiliary field formulation, which better tracks the averaged oscillations of the axion field without requiring a late-time transition.

In the following section, we therefore adopt $m / H_* = 50$ and \texttt{accuracy=1} as a practical compromise between computational efficiency and accuracy, ensuring the CMB power spectra remain near ground-truth levels across a wide range of axion masses.

\section{Axion Constraints} \label{sec:constraints}

We perform a Markov Chain Monte Carlo (MCMC) analysis of the axion model, using the public \textsc{Cobaya}~\citep{Torrado:2020dgo} code. We focus on the mass range $10^{-28}\,\mathrm{eV}\lesssim m\lesssim 10^{-24}\,\mathrm{eV}$, where the standard EFA can introduce significant systematic errors compared to the PH EFA.

We assume a spatially flat cosmology, and adopt flat priors for the six base cosmological parameters:
\[
  \{\,H_0,\, \Omega_\mathrm{D} h^2,\, \Omega_\mathrm{b} h^2,\, n_\mathrm{s},\, \log(10^{10} A_\mathrm{s}),\, \tau\},
   \,.
\]
Here, $\Omega_\mathrm{b} h^2$ is the baryon density, $n_\mathrm{s}$ is the spectral index, $\log(10^{10} A_\mathrm{s})$ is the log of the primordial power spectrum amplitude, and $\tau$ is the optical depth to reionization. We also use a flat prior for the axion fraction,
\[
0 \;\le\; f_\text{ax} \;\le\; 1.
\]

Following Ref.~\cite{Rogers:2023ezo}, we do not fully sample over the axion mass, as the joint posterior in the two-dimensional $(m,f_\text{ax})$ space can exhibit multimodality and other complexities. Instead, we perform separate MCMC runs at fixed discrete values of the axion mass, specifically $m\in\{10^{-28},10^{-27},10^{-26},10^{-25},10^{-24}\}\,\mathrm{eV}$. This strategy captures the dependence of cosmological observables on each chosen mass without requiring a global fit across the entire mass range. Because our main interest is in how constraints on the axion fraction vary with mass, analyzing each mass independently is sufficient to map out the relevant parameter space. 

For each run we use the standard Metropolis-Hastings method to sample over the parameter space, and run multiple MCMC chains in parallel to assess convergence. The convergence criterion is based on the Gelman-Rubin statistic $R_1$, which we require to be $R_1 \;\le\; 0.05$ for each sampled parameter before terminating the chains. This threshold ensures that any remaining variation among chains is negligible, indicating robust sampling of the posterior distribution. After confirming $R_1 \le 0.05$, we discard an initial burn-in portion of each chain and combine the remaining samples to form the final posterior. 

Once the MCMC chains are converged, we perform a BOBYQA minimization of the $\chi^2$ function, using the best-fit point from the MCMC as our initial guess. This step refines the maximum-likelihood estimate, but primary inferences on cosmological parameters and axion properties are derived from the posterior distributions.

Our analysis incorporates \Planck\ CMB power spectra from the PR4 (NPIPE) release, utilizing the latest \textsc{CamSpec} likelihood~\cite{rosenberg22}, along with BAO measurements from DESI~\cite{DESI:2024mwx}. The PR4 release provides approximately 10\% tighter constraints on cosmological parameters compared to the 2018 PR3 release, thanks to reduced noise in the calibrated NPIPE frequency maps. Combining \Planck\ with DESI BAO measurements yields a comparable improvement in constraints on baseline cosmological parameters, relative to earlier BAO datasets used in the 2018 \Planck\ analysis. Consequently, we expect at least part of the improvement in our results over previous studies to stem from this new data.

\begin{table}[t]
    \centering
    \renewcommand{\arraystretch}{1.5}
    \begin{tabular*}{\columnwidth}{@{\extracolsep{\fill}} l c c}
    \hline\hline
    $m$ & $f_\text{ax}$ & $\Omega_{\mathrm{ax}}\,h^2$ \\
    \hline
    $10^{-24}\,\mathrm{eV}$ & $<0.94$ & $<0.110$ \\
    $10^{-25}\,\mathrm{eV}$ & $<0.59$ & $<0.070$  \\
    $10^{-26}\,\mathrm{eV}$ & $<0.044$ & $<0.0051$  \\
    $10^{-27}\,\mathrm{eV}$ & $<0.021$ & $<0.0025$  \\
    $10^{-28}\,\mathrm{eV}$ & $<0.0082$ & $<0.0010$  \\[0.4em]
    \hline
    \end{tabular*}
    \caption{$2\sigma$ upper limits on the axion fraction, $f_\text{ax}$, and physical density, 
    $\Omega_{\mathrm{ax}}\,h^2$, at fixed masses $m$, derived from a joint 
    analysis of \textit{Planck} PR4 and DESI BAO data.}
    \label{tab:axion-constraints}
\end{table}

The constraints on $\Omega_{\mathrm{ax}}\,h^2$ and $f_\text{ax}$ are presented in Fig.~\ref{fig:m_axion_vs_f_axion} and Table~\ref{tab:axion-constraints}. Our analysis has improved constraints on $\Omega_{\mathrm{ax}}\,h^2$ compared to those of Ref.~\cite{Rogers:2023ezo}, which relied on the standard EFA and \Planck\ DR3 data (we obtain approximately 50\% tighter limits for the case $m = 10^{-28}\,\mathrm{eV}$). Ref.~\cite{Rogers:2023ezo} also incorporated DR3 data with full-shape galaxy power spectrum and bispectrum measurements from the Baryon Oscillation Spectroscopic Survey (BOSS). This combination provides stronger constraints by accessing higher-wavenumber modes than the CMB. Our results are comparable to these combined analyses, except for the case of $m = 10^{-25}\,\mathrm{eV}$, where the inclusion of BOSS data yields tighter constraints (by approximately 50\%).

\section{Conclusions}

In this work, we have presented a numerically efficient and accurate implementation of the Passaglia-Hu effective fluid approximation (PH EFA) for ultralight axions (ULAs) within the cosmological Boltzmann solver \texttt{CAMB}.  Our implementation addresses key limitations of the standard EFA, which can introduce significant systematic errors, particularly when modelling axions with masses where oscillations begin around matter-radiation equality, roughly $10^{-27}$ to $10^{-28}$ eV.  By employing auxiliary fields, the PH EFA effectively removes rapid oscillations, allowing for a more accurate representation of the axion field's behavior, even when switching from the Klein-Gordon equation to the fluid approximation at relatively early times.

We demonstrated the robustness of the PH EFA, showing that, compared to the standard EFA, it requires much smaller variations in the initial field value to match a target present-day axion density. Together with the improved modelling of axion perturbations, a comparison of the mismatch measure between model and ground-truth CMB spectra highlights the improved accuracy of the PH EFA. It also reduces the need for fine-tuning the switch time, a nuisance parameter in the standard EFA.

Furthermore, we performed a Markov Chain Monte Carlo (MCMC) analysis incorporating our new implementation, using the \Planck\ PR4 CMB data and DESI BAO data.  We obtained improved constraints on the axion mass and abundance, compared to previous studies which use the standard EFA and older datasets. In particular, we find constraints comparable to those found using large-scale structure (LSS) in previous analysis.  Our implementation, \texttt{AxiCAMB}, is publicly available at \url{https://github.com/adammoss/AxiCAMB}, and is based on the latest version of \texttt{CAMB}.

Current CMB and LSS data probes the axion fraction at the percent level for $10^{-30} \, {\rm eV} \lesssim m_{\text{ax}} \lesssim 10^{-28}$ eV. Future experiments will offer even tighter constraints on the ULA parameter space: CMB-S4 will extend sensitivity up to $m_{\text{ax}} \sim 10^{-24}$ eV \cite{Hlozek:2016lzm}, while intensity mapping from the Square Kilometre Array will be sensitive to $f_{\text{ax}} \sim {\cal O}(0.01-0.001)$ for $10^{-32} \, {\rm eV} \lesssim m_{\text{ax}} \lesssim 10^{-23}$ \cite{Bauer:2020zsj}. This highlights the need for highly accurate numerical methods, such as the one presented here, for exploiting future data.

We highlight several avenues where our approach could be extended. First, while we have focused on ULAs with a simple quadratic potential, axion-like fields with general potentials (including full cosine forms) or self-interactions may lead to new signatures; generalizing the auxiliary-field method to these scenarios is a natural next step. Second, 
the synergies with next-generation probes such as CMB-S4, the Simons Observatory, and the SKA will significantly improve limits on sub-horizon physics, and further investigation is warranted to examine the robustness of the method across broader parameter ranges, including the regime of heavy axions. Finally, the axion sector may include multiple light scalars, and extending the PH formulation to such multi-field models could clarify whether ULAs can address cosmological tensions and open new windows on the dark sector.

\textit{Note added:} As we were nearing completion of this paper, Ref.~\cite{Liu:2024yne} appeared with a similar implementation of the PH EFA, based on the original \textsc{AxionCAMB} codebase. We have compared our results with theirs in the regime of sufficiently large switch times, $m/H_* \gtrsim 30$, and find very good agreement in the resulting CMB power spectrum.

\section*{Acknowledgments}

The work of AG and AM was supported by an STFC Consolidated Grant [Grant No. ST/T000732/1]. LG was supported by an STFC studentship [Grant No.\ ST/X508639/1]. For the purpose of open access, the authors have applied a CC BY public copyright license to any Author Accepted Manuscript version arising.

\bibliographystyle{apsrev4-2}
\bibliography{refs}

\end{document}